# Visualising Virtual Communities: From Erdős to the Arts

Jonathan P. Bowen
London South Bank University
Department of Informatics
Faculty of Business
Borough Road, London SE1 0AA
United Kingdom
jonathan.bowen@lsbu.ac.uk
http://www.jpbowen.com

Robin J. Wilson
Pembroke College
University of Oxford
Oxford OX1 1DW
United Kingdom
r.j.wilson@open.ac.uk
http://www.pmb.ox.ac.uk/Fellows_Staff/?profile=259

**Monitoring communities has become increasingly easy on the web as the number of visualisation tools and amount of data available about communities increase. It is possible to visualise connections on social and professional networks such as Facebook in the form of mathematical graphs. It is also possible to visualise connections between authors of papers. In particular, Microsoft Academic Search now has a large corpus of information on publications, together with author and citation information, that can be visualised in a number of ways. In mathematical circles, the concept of the "Erdős number" has been introduced, in honour of the Hungarian mathematician Paul Erdős, measuring the "collaborative distance" of a person away from Erdős through links by co-author. Similar metrics have been proposed in other fields, including acting. The possibility of exploring and visualising such links in arts fields is proposed in this paper.**

*Virtual community. Visualisation tools. Academic search. Arts community. Mathematical graphs.*

## 1. INTRODUCTION

During the past decade, the web has been adapting to provide an increasing sense of community, including in cultural and museum spheres (Bowen 2000, Beler *et al.* 2004). Social networking and collaborative features provide the possibility of feedback and interaction between web users and institutions (Borda & Bowen 2011, Liu & Bowen 2011). For example, blog-style interaction currently available on the web allows connections between people with an interest in the arts (Beazley *et al.* 2010, Liu *et al.* 2010).

A community of people can be modelled naturally as a mathematical graph (Aldous & Wilson 2000), as can the web itself with its pages and hyperlinks (Numerico *et al.* 2004). The people can be represented as vertices in the graph and links between them as edges of the graph. These edges may be undirected (e.g., for friendship between two people where both like each other, or as co-authors of a joint publication) or directed (e.g., for a citation of one author to another author's paper). Such a graph is a natural way to visualise relationships between people. When observed visually, patterns in the graph can be quickly assimilated by the viewer.

With the increase in social and professional networking online, the visualisation of communities in an automated way as graphs is now relatively easy. For example, Figure 1 shows connections between one of the authors and "friends" on Facebook, using visualisation software provided by TouchGraph (http://www.touchgraph.com). The tool also takes account of links between all the people included in the graph. Thus it is possible to note groups within the network visually. Greatly interconnected groups of people are clustered together physically on the displayed graph and are highlighted using colours. For example, in the case of Figure 1, those marked in light blue are mainly people interested in computer science and those marked in pink are mainly interested in museums and the arts, two major but largely non-overlapping areas of interest to the first author.

In this paper, we mainly consider the visualisation of academic communities. This is currently easier in the sciences, due to the availability of information on academic publications online, but for the future it will be increasingly possible in the arts and humanities as well. We consider academic communities through co-authorship, especially as recorded in the online Academic Search database, which has good visualisation facilities.

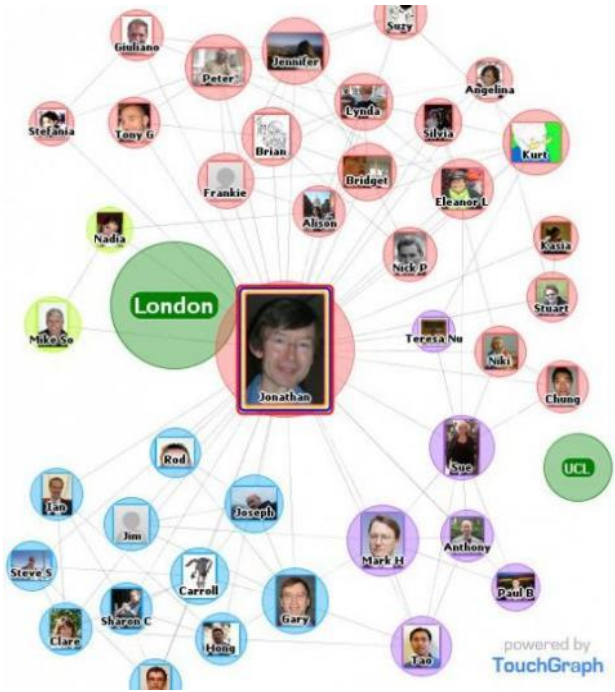

*Figure 1.* Facebook TouchGraph connections

## 2. ACADEMIC COMMUNITIES

Academic links are often indicated by co-authorship of publications and by citation to other authors in papers. Databases holding such information are available. Traditionally these have been by subscription, and typically for university use, although now there are generally available online databases that do not require payment for their use. In this paper we concentrate on the latter. In any case, the interface normally provides a mainly textual view with search facilities that return a list of the papers that have been found. Typically there are no visualisation facilities to display relationships between the items in the database.

In computer science, an increasing number of resources on publications are available online (Bowen 2011). Existing databases, such as those from professional societies like the ACM Digital Library (http://dl.acm.org) and the IEEE Xplore Digital Library (http://ieeexplore.ieee.org), as well as independent resources like the DBLP Computer Science Bibliography (http://dblp.uni-trier.de), cover computer science in general, providing a standard web interface with search facilities and largely textual output. Other disciplines are often less well covered online, especially in the arts and humanities.

Now web-based search providers like Google and Microsoft are developing databases of publications with interactive facilities for updates. These cover a wider range of fields, still biased towards the sciences, but also with the potential to cover arts and humanities in a more comprehensive manner (Microsoft Academic Search 2012).

Google Scholar (http://scholar.google.com) has provided a very extensive database of academic publications for a while and more recently has added a feature that allows authors to take control of their own publications, enabling these to be presented as a corpus of work. The facility includes the ability to include keywords that allows grouping of authors, although there is no control of these keywords or linking of similar terms. For example, there are some authors who have included "visualization" and others who have used the term "visualisation", with no connection between the two.

For individual authors, a graph of citations per year is included, along with the total number of citations, the author's *h-index* (the number of top *h* papers each with at least *h* citations) and *i10-index* (the number of papers with at least ten citations), both for all years and for the last five years (see Figure 2). It is possible to add links to co-authors that also have an individual entry on Google Scholar. To use this facility, it is necessary to register with Google and log into Google Scholar. A convenient "My Citations" link is included at the top right of the main Google Scholar page.

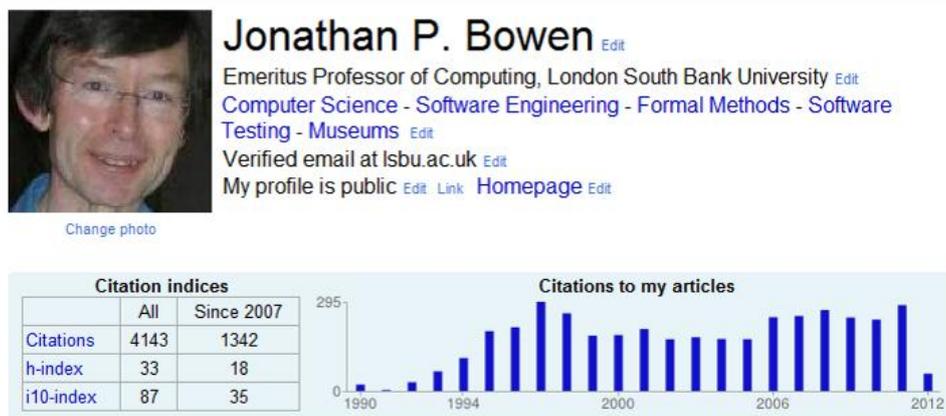

*Figure 2.* Google Scholar personal page.

Microsoft Academic Research from Microsoft Research (http://academic.research.microsoft.com) in Beijing provides a database of publications across many different disciplines. This has better visualisation features than Google Scholar, but a less extensive corpus of publications. There is a different approach to Google Scholar, where only an individual's publications can be edited. On Microsoft Academic Search, any entry can be edited, but it is checked before it is actually updated for viewing (a process that can take from days to weeks, depending on the backlog and the effort provided by Microsoft).

Google Scholar is better at merging publications by a given author correctly, but Academic Search allows manual merging of authors. There is an "Arts & Humanities" section on Academic Search which collects together authors, conferences, journals, etc., that are relevant to the field (Microsoft Academic Search 2012). Entries for individual authors include a graph of papers and citations by year, together with the author's *h-index* (see above) and *g-index* (the number of top *g* articles with a total of at least $g^2$ citations). It is possible to view co-authors, citing authors, and transitive links between pairs of authors through co-authors, as well as supervisor/student relation-ships, in a visual manner. To update information on Academic Search, it is necessary to register and log in with a Windows Live account.

### 3. ACADEMIC SEARCH

In this section we provide more detailed information on the visualisation facilities provided by Microsoft Academic Search since, in our experience, this is the best of such facilities that is freely available online.

Academic Search offers a number of visualisation features. On each author page (see Figure 3) there is a graph combining both the number of papers (in pale blue) and the number of citations (in orange) by year on the same axes, each with appropriate scales. Initially by default, this is displayed in accumulative mode, showing the total number of papers/citations up to that point for each year. A more useful annual display of the totals for each individual year can also be selected. This gives a course view of the productivity of the author and also the influence of the author over their career so far, although a note of caution should be made concerning such crude metrics (Parnas 2007). It should also be noted that the database is not entirely accurate, since much of the information has been generated automatically without human intervention. Gradually mistakes in the Academic Search database are being eliminated through human checking and correction by online users with an interest in academic publications.

Typically there is a lag of a few years for the citations, compared to the publications, as would be expected. There is also a rapid dropping off for the last couple of years, since the database is not completely up to date.

As well as information on individual authors, visualisations are also available showing the connections of co-authors, citing authors, super-visor/student trees, and transitive links between any two authors within the database.

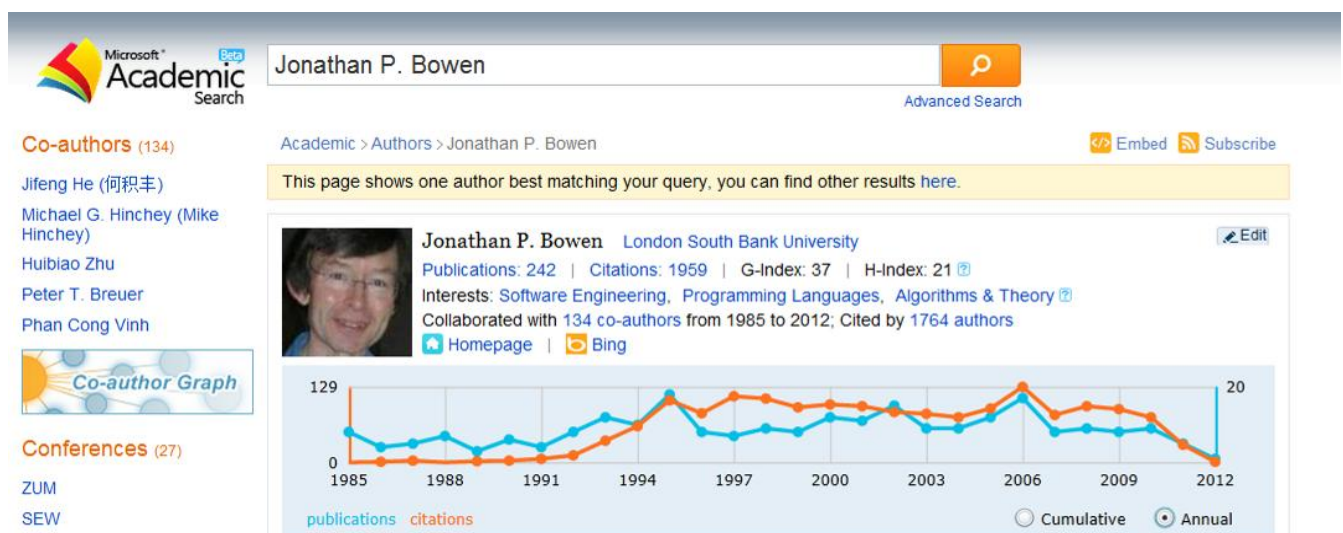

*Figure 3.* Academic Search personal page.

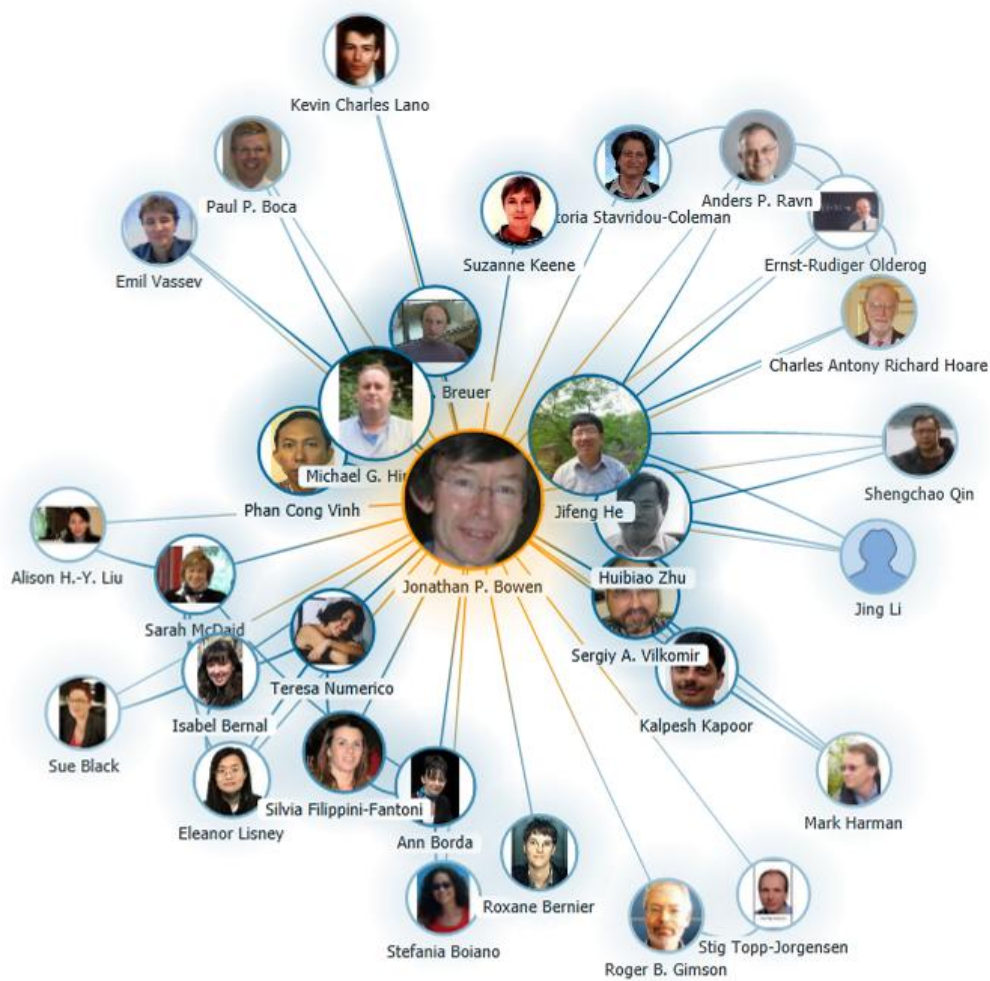

*Figure 4.* Academic Search co-author graph.

The view of co-authors (see Figure 4) is by default restricted to the top 30 co-authors as measured by the number of publications that have been written together. The limited number helps to avoid the view from becoming too cluttered for prolific authors with a large number of co-authors. Co-authors with the most joint publications are shown closer to the main author. Links are also shown between pairs of co-authors as well, so clusters can appear where a group of co-authors write a number of publications together. If the visualisation is too cluttered in any particular area, it is possible for the viewer to magnify the view or even move around individual co-authors using the mouse, with the connecting lines moving appropriately.

Initially the main author is shown in the centre, with co-authors distributed 360 degrees around that author, who is also displayed larger that the other co-authors. Each author is identified by a small circular thumbnail photograph if available (or a standard silhouette image if not) together with their name below. Authors with a higher number of co-written papers are displayed with a larger image as well is being closer to the main author.

In Figure 4, clustering of co-authors can be observed. Those to the top right are largely in the field of software engineering (the main research area of this illustrated author) whereas those to the bottom left are in the field of museum informatics.

Citing authors can be displayed in a similar way to co-authors (see Figure 5), although in this case the cited author is displayed at the top left initially, with the citing authors displayed in a 90-degree quadrant to the right and downwards. Authors who have cited the main author most are displayed closest and with a larger image.

Supervisors and their doctoral students can be listed together in the form of a tree. See Figure 6 for a tree associated with the mathematician Alonso Church, including one of his most notable students, Alan Turing. Where large numbers of students are involved, these are collected together by their institution, as in Figure 6. This part of the Academic Search database is not very well populated at the moment. To update the information, evidence is required via a web link provided by the person requesting the update (e.g., to a university web page with suitable information).

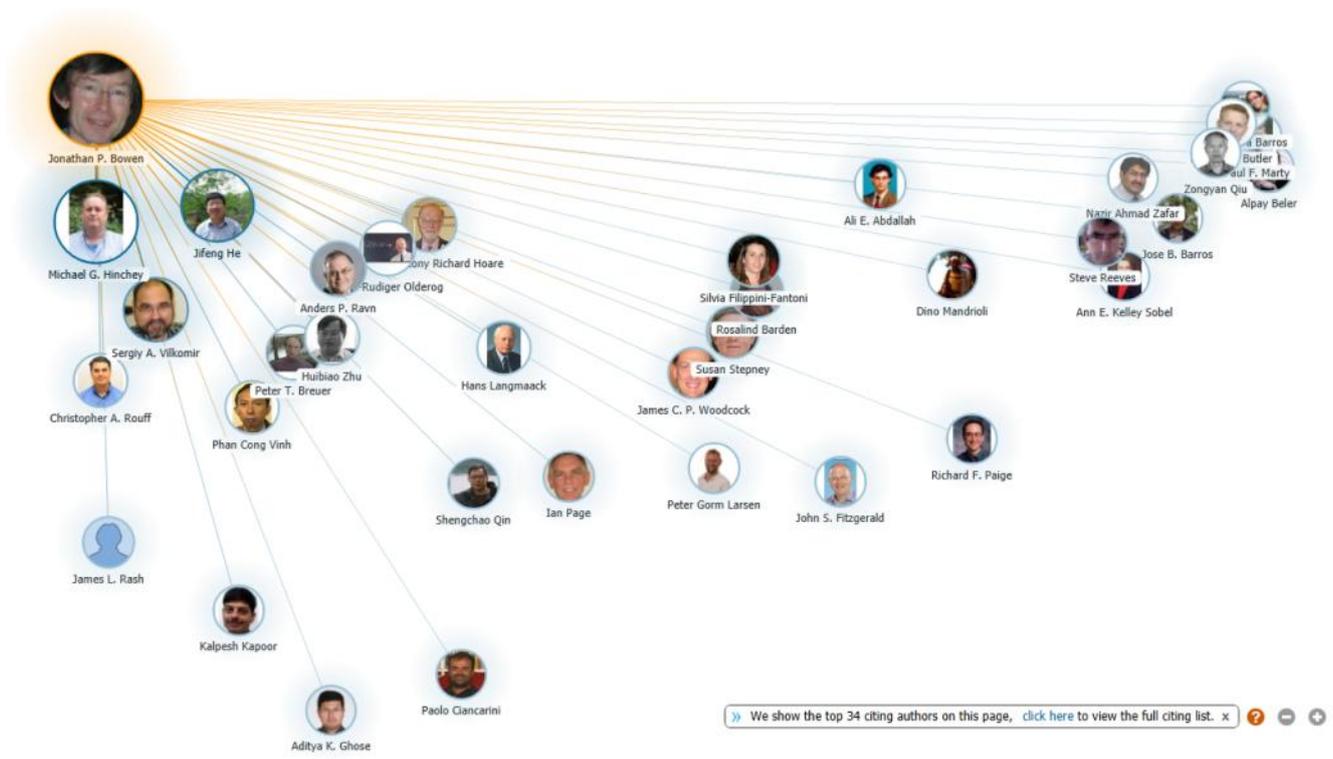

*Figure 5. Academic Search citation graph.*

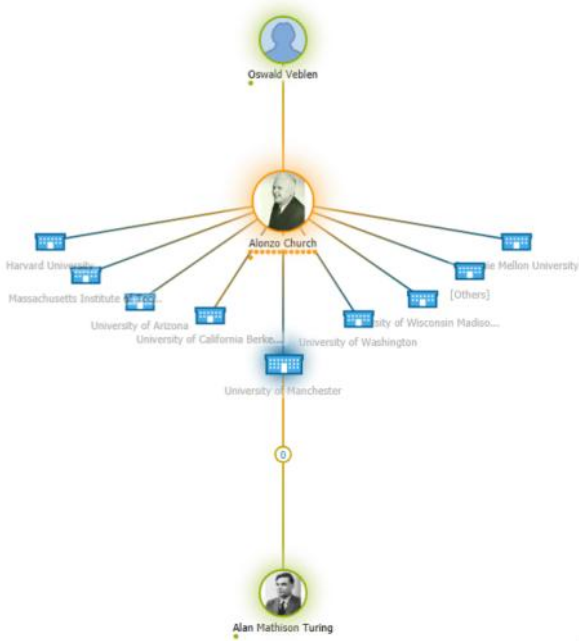

*Figure 6. Academic Search genealogy graph.*

An interesting and flexible feature of Academic Search is the option to provide two authors and the links between them to be displayed transitively by co-authorship of papers (see Figure 7). For two co-authors, this would simply be a link between those two authors. For authors who have not co-written any papers, the software searches for a selection of the most direct links between the two authors via co-authors. For example, if another author had written at least one paper with the two authors in question, a connection with that co-author linked between the two authors in question would be displayed. By default, Academic Search provides a set of links between the current author and the mathematician Paul Erdős, although any author can be selected.

## 4. THE ERDŐS NUMBER

The Hungarian mathematician Paul Erdős (1913–1996) (http://en.wikipedia.org/wiki/Paul_Erdős) co-authored over 1,000 publications, collaborating with 511 people during his lifetime. This has inspired the concept of the "Erdős number" within mathematical circles (http://en.wikipedia.org/wiki/Erdős_number).

Paul Erdős himself was the only person with an Erdős number of 0. All his co-authors have an Erdős number of 1, including the second co-author of this paper (Erdős & Wilson 1977). Figure 7 shows a direct link between Robin Wilson and Paul Erdős due to their joint paper published in 1977, as well as a number of other less direct transitive links through other co-authors. As a result of co-writing this current paper, the first co-author now has an Erdős number of 2 and all co-authors of the first author have an Erdős number of at most 3.

The Erdős Number Project at Oakland University (http://www.oakland.edu/enp/compute/) studies the research collaboration between mathematicians. It provides help in computing the Erdős number for mathematical authors as well as examples of famous scientists with known Erdős numbers.

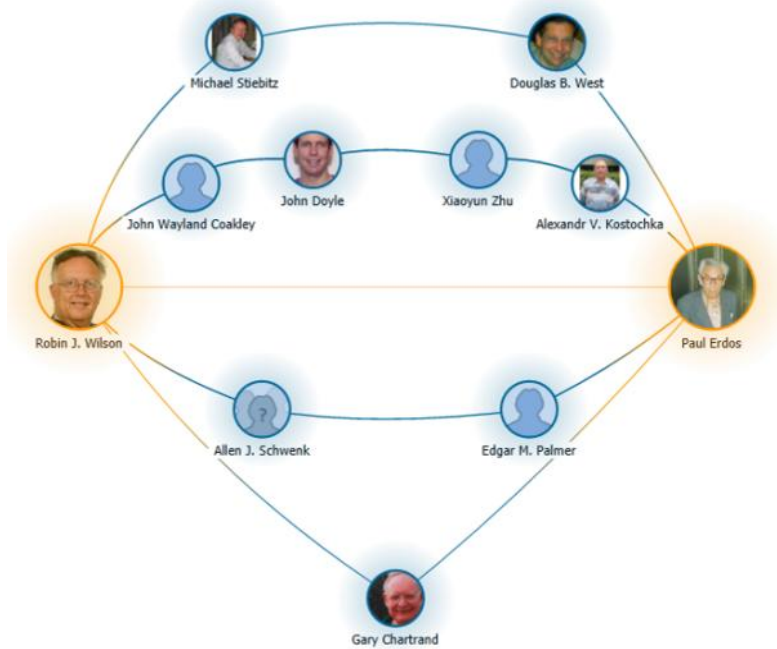

*Figure 7.* Academic Search co-author path

As has been noted elsewhere, the entire population of the world is connected by six degrees of separation or fewer through personal connections (http://en.wikipedia.org/wiki/Six_degrees_of_separation). In the case of the Erdős number, the degree of separation for people with any significant relationship to mathematical publishing is even less, typically 3 or 4.

In the arts, specifically in the world of film, a similar "Bacon number" has been developed, named after the film and theatre actor Kevin Bacon (born 1958, http://en.wikipedia.org/wiki/Bacon_number). This is like the Erdős number, but applies to film credits instead of publications.

Provocatively, the idea of a combined measure, the "Erdős–Bacon number", has also been proposed (http://en.wikipedia.org/wiki/Erdős–Bacon_number). This is the sum of a person's Erdős and Bacon numbers. Due to the interconnectivity between the mathematical and arts worlds, there are both scientists and actors with surprisingly low Erdős–Bacon numbers, as small as 3 for example, and a significant number below 10.

Further such numbers could be suggested for arts and humanities fields, using a prolific collaborator as a starting point.

**5. CONCLUSION**

Visualising virtual communities has become increasingly easy over the past decade as social and professional networking has developed rapidly. There are even specific online communities for academics. For example, a leading example is Academia.edu (http://academia.edu), including facilities for adding papers together with metadata, monitoring access statistics, and developing contacts with other users. Data on academic connections is now readily available online, especially in scientific circles, but increasingly in the arts and humanities as well. This paper has given some examples of what is possible at the moment.

For the future, it is expected that visualisation of arts-related communities will improve to be on a par with that available for the scientific community, e.g., through the addition of the necessary data to existing databases or the addition of visualisation facilities to the interfaces for arts databases. For example, the Internet Movie Database (IMDb, http://www.imdb.com) online film resource is very extensive, but does not currently provide any visualisation facilities, as available on Academic Search for example.

Communities of practice, as postulated in the social science field, could be studied further, both in scientific (Bowen & Reeves 2011) and arts-related communities (Liu *et al.* 2010). In particular, visualisation of these communities dynamically over time could help in understanding their nature as they grown and contract.

In summary, it can be expected that visualisation of online communities will improve significantly over the next decade just as the communities themselves have developed and expanded rapidly over the past decade. Arts resources have the potential to provide a rich resource for such visualisation, especially in the case of collaborative arts endeavours where multiple artists are involved.

**Acknowledgements**

Jonathan Bowen thanks Museophile Limited (http://www.museophile.com) and Ingrid Beazley for support in attending the EVA London 2012 conference.